
\input phyzzx
\rightline{EFI-93-16}
\title{\bf Matrix Models and Nonperturbative String Propagation
\break in Two-dimensional Black Hole Backgrounds}
\vglue-.25in
\author{Sumit R. Das \foot{On leave of absence from Tata Institute of
Fundamental Research, Bombay 400005, INDIA} \foot{e-mail:
sudas@yukawa.uchicago.edu}}
\address{Enrico Fermi Institute
\break University of Chicago
\break Chicago, Illinois 60637.}
\vglue-.25in
\abstract{We identify a quantity in the $c=1$ matrix model which
describes the wavefunction for
physical scattering of a tachyon from a black hole of the
two dimensional critical string theory. At the semiclassical level
this quantity corresponds to the usual picture of a wave coming in
from infinity, part of which enters the black hole becoming
singular at the singularity, while the rest is
scattered back to infinity, with nothing emerging from the whitehole.
We find, however, that the exact nonperturbative wavefunction is
nonsingular at the singularity and appears to end up
in the asymptotic region ``behind'' the singularity.}
\endpage
\def\tV{{\tilde{V}}}
\def\tW{{\tilde{W}}}
\def\paz{\partial_z}
\def\half{{1 \over 2}}
\def\pax{\partial_x}

\def\psd{\psi^{\dagger}}
\def\smu{{\sqrt{\mu}}}
\def\stwo{{\sqrt{2}}}
\def\tT{{\tilde  T}}
\def\twa{{2 \over \alpha '}}

\def\pax{\partial_x}
\def\half{{1 \over 2}}
\def\pat{\partial_t}

\def\pap{\partial_p}
\def\smu{{\sqrt{\mu}}}

\def\twa{{2 \over \alpha '}}

\def\rega{{\rm Region~I}}
\def\regb{{\rm Region~II}}

\def\rtr{{\rm Tr}}
\def\tX{\tilde X}
\def\tS{\tilde S}

In a previous paper \Ref\SD{S.R. Das, Mod. Phys. Lett. A8 (1993) 69}
it was suggested that the one dimensional matrix model
\Ref\DONEG{D. Gross and N. Milkovic, Phys. Lett. 238B (1990) 217; E.
Brezin, V. Kazakov and Al. B. Zamolodchikov, Nucl. Phys. B338 (1990) 673;
G. Parisi, Phys. Lett. 238B (1990) 209; P. Ginsparg and J. Zinn-Justin,
Phys. Lett. 240B (1990) 333.}
may be used to study the nonperturbative behavior of
massless ``tachyons'' coupled to the black hole
background of the two dimensional critical string
\Ref\BLACKA{G. Mandal, A. Sengupta and S.R. Wadia, Mod. Phys. Lett. A6
(1991) 1685}
\Ref\BLACKB{E. Witten, Phys. Rev. D44 (1991) 314.}. This
follows from the fact that at the semiclassical level
a certain integral transform of the
fluctuation of the
collective field of the matrix model \Ref\JSAK{ A. Jevicki
and B. Sakita, Nuclear Physics B165 (1980) 511} \Ref\DJEV {S.R. Das and
A. Jevicki, Mod. Phys. Lett. A5 (1990) 1639} around the classical
ground state value satisfies the same
{\it linearized} equation of motion as that of the massless tachyon
in a blackhole background.
Indeed this connection between the matrix model and the black hole  is
essentially the same as that between
the liouville theory and the $SL(2,R)/U(1)$ coset model discovered earlier
by Martinec and Shatashvilli \Ref\MART{
E. Martinec and S. Shatashvilli,
Nucl. Phys. B368 (1992) 338.}. It was soon found in \Ref\DMWBH{A.
Dhar, G. Mandal and S. Wadia, Mod. Phys. Lett A7 (1992) 3703.} that a similar
transformation exists in the bosonized form of the fermionic
field theory of the matrix model developed in \Ref\DMW{A. Dhar, G. Mandal
and S. Wadia, Int. J. Mod. Phys. A8 (1993) 325; Mod. Phys. Lett.
A7 (1992) 3129.} and leads to the same black hole interpretation. A large
class of transformations between collective field theory and black hole
wavefunctions have been
discussed in \Ref\JY{A. Jevicki and T. Yoneya, in T. Yoneya, U. of Tokyo
(Komaba) Preprint UT-KOMABA-92-13, hepth/9211079.}. In fact, there
are several pieces of evidence from continuum treatments that the liouville
and black hole backgrounds are closely related
 \Ref\CONT{M. Bershadsky and D. Kutasov, Phys. Lett. B266 (1991) 345;
T. Eguchi, H. Kanno and S.K. Yang, Phys. Lett. B298 (1993) 73;
S. Mukhi and C. Vafa, Harvard and TIFR Preprint HUTP-93/A002,
TIFR/TH/93-01, hepth 9301083}.

The presence of this connection does not imply that the physics of
the black hole background is identical to that of the liouville
background. Rather, at the semiclassical level,
the integral transform allows one to obtain the behaviour of a
tachyon in a black hole background from matrix model quantities just
as singular gauge transformations in gauge theories allow one to obtain
wavefunctions in presence of vortices or monopoles from those in free
space.
However, if one examines the structure of the integral transform,
one finds that near the singularity of the black hole
this receives contributions from the collective
field in regions where the coupling is strong.
Far from the black hole the semiclassical approximation is good so
that one can meaningfully talk about incoming and outgoing tachyon
states.

Beyond the semiclassical expansion the matrix model quantities transformed
as above would represent correlations in {\it some} quantum field
theory of scalar particles whose semiclassical behaviour is identical
to that of tachyons moving in a static black hole. We could thus use
matrix model results to study non-perturbative behaviour of tachyon
propagation.
 Indeed one feature of such non-perturbative effects
have been observed in \refmark\DMWBH where it was shown that, with the
particular definition of this transform used in that paper, the
classical value of the {\it background} is singular at the location of the
black hole singularity, but the exact value is completely regular.
A similar behavior has been found for the transform of a class of wavepackets
which represent ``spikes'' on the fermi sea \Ref\DMWDAE{A. Dhar, G. Mandal
and S.R. Wadia - talk by G. Mandal in DAE Symposium on High Energy Physics,
Bombay, December 1992; see also
TIFR preprint TIFR-TH-92-40A, hepth/9212027}.
However these spikes transform into tachyon
wavefunctions which do not represent scattering processes as described
above.

What is lacking in all the papers quoted above is a clear identification
of the quantities which have to be computed in the {\it matrix model} which
represent a {\it physical} scattering process in the black hole background.
For example, at the semiclassical level, the transforms defined in either
\refmark\SD or \refmark\DMWBH represent a process where there is a
non-zero flux emerging from the past null infinity and hence do not
represent a process in which the scalar particle comes in from infinity,
part of the wave being scattered and part of the wave absorbed by the
black hole. One of the transforms considered in \refmark\JY does represent
a scattering type wavefunction in the exterior region, but not in the
interior region - where the wave has a component which does not vanish on
the extension of the past horizon.

In this paper we make this identification by noting that the large
class of transforms which take matrix model quantities into single
particle wavefunctions in a black hole background correspond to
different boundary conditions \refmark\MART \refmark\JY.
We find the correct linear combination
of these transforms
which corresponds to a wave vanishing on the
past horizon and hence represents a physical scattering process
from a black hole. We show that, as expected,
single particle wave functions which represent such a scattering process
diverge at the black hole singularity at the semiclassical level.
We then
consider quantum corrections to this single particle wave function using
the exact results of the matrix model. We show that while to all orders
of string perturbation theory the divergence of the wavefunction at the
location of the singularity persists, the exact non-perturbative answer
is completely nonsingular. We show that in the aymptotic region
``beyond'' the singularity this wave corresponds to a purely outgoing
wave. Thus at the exact quantum level the particle thrown in from the
standard asymptotic region in the exterior of the black hole ends up
in a different asymptotic region beyond the black hole.

\noindent{\underbar{The semiclassical limit}} \nextline

Consider the two types of loop operators in the matrix model of
a matrix $M_{ij}(t)$
The macroscopic loop operator with real loop lengths is defined
by the quantity
$$ W(p,t) \equiv \rtr~e^{-pM} =
\int_{2 \smu}^{\infty} dx~e^{-px}~\phi(x,t) \eqn\two$$
We will also define macroscopic loop operators with
{\it imaginary} loop lengths which are defined by
$$ V(z,t) \equiv \rtr~e^{izM} =
\int_{-\infty}^{\infty} dx~e^{izx}~\phi (x,t) \eqn\three$$
The field $\phi(x,t)$ is the density of eigenvalues of $M_{ij}$, the
collective field.
In terms of the fermionic fields $\psi (x,t)$ of the fermionic
field theory of
\Ref\FERM{A. Sengupta and S.R. Wadia,
Int. J. Mod. Phys. A6 (1991) 1961; D. Gross and I. Klebanov, Nucl. Phys.
B352 (1990) 671}
one has $\phi (x,t) = \psd (x,t)
\psi (x,t)$.
In the fermionic field theory the correlations of $W(p,t)$ may be
obtained by first computing the correlations of $V(z,t)$ and then
performing a suitable analytic continuation, as explained in
\Ref\MORE{G. Moore, Nucl. Phys. B368 (1992) 557.}.  However, in the
considerations which follow it is important to keep in mind the
difference between the two quantities.

In the lowest order of the semiclassical expansion one can obtain
linearized equations for the matrix elements of the loop operators
between one particle states and the ground state, which
we will denote by $\tV$ and $\tW$. Semiclassically these may be
obtained by expanding the collective field $\phi(x,t)$ around its classical
ground state value $\phi_0(x,t) = {\stwo \over \pi}(x^2 - 4 \mu)^{\half}$
$$ \phi (x,t) = \phi_0 (x,t) + \pax \eta (x,t) \eqn\five$$
and replace $\phi(x,t)$ in \two\ and \three\ by $\pax \eta(x,t)$.
Using the equations of collective field theory it may be seen that at the
classical level the linearized equation of motion for $\eta(x,t)$ leads
to the following linearized equation for $\tW(p,t)$ and $\tV(z,t)$
$$ \eqalign{&[(p \pap)^2  - \pat^2 - 4 \mu p^2] \tW(p,t) = 0 \cr
&[(z \paz)^2  - \pat^2 + 4 \mu z^2] \tV(z,t) = 0}\eqn\fivea$$
The first equation in \fivea\ is identical to the Wheeler-de Witt
equation obtained in the liouville theory. Indeed using the
semiclassical expression for energy eigenstates
 $\eta (x,t)= e^{i\omega t}~sin(\omega~cosh^{-1}({x \over 2 \smu}))$
of the matrix model satisfying
the correct Dirichlet boundary condition the fluctuation of $W(p,t)$
is given by the modified Bessel function $e^{i\nu t}~
K_{i \nu}(2 \smu~p)$ which represents (in the coordinate $log~p$) a
wave coming in from infinity and getting reflected perfectly from the
``liouville barrier''. This is of course the basic picture of scattering
in the liouville theory. Formally $\tV(z,t)$ may be regarded as the
liouville wave function with {\it negative} cosmological constant. We
shall not, however, assign any physical meaning to $\tV(z,t)$ and we
will not need any.

Consider now the following classes of transforms of $\tV$ and $\tW$
$$\eqalign{&\tT_{\pm}^{\pm}(u,v) \equiv \int_0^\infty dz
\int_{-\infty}^{\infty}
dt~e^{\pm iz(e^t v - e^{-t}u)}~\tV(\pm z,t) \cr &
\tS_{\pm}(u,v) \equiv \int_0^\infty dp \int_{-\infty}^{\infty}
dt~e^{\pm iz(e^t v + e^{-t}u)}~\tW(p,t)} \eqn\xten$$
Using equations \fivea\ it may be easily verified that each of these
quantities $\tT^\pm_\pm, \tS_\pm$ satisfy the following linearized
equation
$$[4(uv + \mu) \partial_u \partial_v + 2(u\partial_u + v\partial_v) + 1]
\tX_i (u,v) = 0 \eqn\six$$
where $\tX_i$ stands for any of the $\tT_\pm^\pm, \tS_\pm$
This is precisely the linearized equation of motion for the massless tachyon
of the two dimensional critical string moving in a black hole background.
The coordinates $(u,v)$ are the Kruskal-like coordinates in terms of which
the background metric and dilaton fields are \foot{Our conventions are
those of \refmark\BLACKA}
$$ \eqalign{G_{uv} = & G_{vu} = {1 \over 2(\twa~uv + a)}~~~G_{uu} = G_{vv}
= 0 \cr & D(u,v) = -\half~{\rm log}~(\twa~uv + a)}\eqn\fourteen$$
The black hole mass is given by $a = \twa \mu$. In terms of the coordinates
$(u,v)$ the various regions of the black hole geometry are
$$ \eqalign{ & u = r~e^\theta~~~~v = r~e^{-\theta}~~~~{\rm for}~~~u,v
\ge 0~~~~{\rm Region~ I} \cr
& u = -r~e^\theta~~~~v = r~e^{-\theta}~~~~{\rm for}~~~u < 0, v > 0
{}~~~~{\rm Region~II} \cr & u = - r~e^\theta~~~~v = - r~e^{-\theta}
{}~~~~{\rm for}~~~u,v < 0 ~~~~ {\rm Region~III} \cr & u = r~e^\theta
{}~~~~v = - r~e^{-\theta}~~~~{\rm for}~~~u > 0, v < 0
{}~~~~ {\rm Region~IV}} \eqn\fivea$$
Thus the Region II contains the future singularity while the Region IV
contains the past singularity.

The transform defined in \refmark\SD is given by $\tS_+$, while the
transform defined in \refmark\DMWBH is a specific linear combination of
all of these independent transforms \foot{In \refmark\DMWBH the
transform involves a collective field which is the density of fermions
in the momentum space conjugate to the eigenvalue coordinate. However,
in the double scaling limit the interchange of momenta and coordinates
simply changes the sign of the cosmological constant. This may be then
used to relate the part of the transform of \refmark\DMWBH involving
momentum space collective fields in terms of the quantities we have
defined above.}. Given a space-time history of
$\phi (x,t)$ these various transforms provide {\it different} space time
histories of a tachyon in a black hole background. Semi classically
these histories may be easily evaluated using the expression for the
fluctuations of the collective field given above and are given by
hypergeometric functions, as done for $\tS_+$ in \refmark\SD.
However, the different transforms lead to different hypergeometric
functions, i.e. waves with different boundary conditions
\Ref\DVV{see R. Dijkgraaf, E. Verlinde and H. Verlinde, Nucl. Phys. B371
(1992) 269.}. For
example $\tS_+$ with $\tW$ chosen to be of the form
$\tW \sim e^{i\nu t}$ is given in Region I by
$$\eqalign{\tS_+(r,\theta) = &{e^{-i\nu \theta} \over r}
[({\mu \over r^2})^{i \nu \over 2}
A(\nu) F( \half - i\nu, \half; 1 - i\nu; -{r^2 \over \mu}) + \cr &
(-1)^{i\nu} ({\mu \over r^2})^{-i \nu \over 2} A(-\nu) F(
\half + i\nu, \half; 1 + i\nu; -{r^2 \over
\mu})]} \eqn\fiveten$$
where $A(\nu) \equiv {\Gamma(i\nu) \over \Gamma(\half + i \nu) \Gamma
(\half)}$ and we have omitted an overall $\nu$-dependent constant.
It is clear from \fiveten\ that near the horizon at $r = 0$ this
represents a linear combination of  a wave coming out of the past
event horizon as well as one going into the future event horizon. This is
not what we want to study for the scattering of a particle from a
black hole. For the latter physical problem we must ensure that there
is nothing emerging from the past horizon.
We specifically
want to study  what happens to the wave at the singularity.
The transform of \refmark\DMWBH also consists of a wave which is
nonvanishing at the past horizon.

It is indeed possible to write down a transform which
represents a physical
scattering of a tachyon from the black hole, i.e. a wave coming
in from infinity - part of which gets inside the horizon while a part is
scattered back with nothing emerging from the past horizon. This is
obtained by first using the time translation invariance of the theory
to work in terms of fourier transforms of the operators $V(z,t)$ defined
by
$$V(z,t) = \int d\nu~e^{i\nu t}~V(z,\nu) \eqn\xeight$$
The correct transform is then
given, in terms of the fourier transformed quantities $\tV(z,\nu)$ by
$$T_{scat}^{(\nu)}(u,v) =  \int_0^\infty dz \int_{-\infty}^{\infty}
dt~[e^{{\pi \nu \over 2}}~e^{iz(e^t v - e^{-t}u)} +~
e^{-{\pi \nu \over 2}}~e^{-iz(e^t v - e^{-t}u)}]~
e^{i\nu t}~[V(z,t) + V(-z,t)]
\eqn\xeleven$$
The above definition may be regarded as an operator definition. To
extract single particle wave functions and hence obtain $\tT_{scat}$
one has to take suitable matrix elements which amounts to the
replacement of $V(z,t)$ by $\tV(z,t)$. The integration over $t$ in
\xeleven\ may be explicitly performed separately in the different
regions, with the result
$$\eqalign{&\tT_{scat}^{(\nu)} = e^{i\nu \theta} \int_0^\infty
dz~[K_{i\nu}(2zr)][\tV(z,\nu) + \tV(-z,\nu)] ~~~~~~\rega \cr &
\tT_{scat}^{(\nu)} = e^{i\nu \theta} \int_0^\infty dz~[H_{i\nu}^{(1)}(2zr)
- H_{i\nu}^{(2)}(2zr)][\tV(z,\nu) + \tV(-z,\nu)]
{}~~~~~~\regb} \eqn\xfourteen$$
The expressions in the other regions may be trivially obtained from these
by changing the sign of one of the coordinates appropriately.

Let us first evaluate this transform semiclassically using
collective field theory as discussed above. This leads to
$$\tV^{sc}(z,\nu) = A(\nu)~H^{(1)}_{i\nu} (2 \smu z)~~~~~~~~
\tV^{sc}(-z,\nu) = -A(\nu)~H^{(2)}_{i\nu} (2 \smu z)
\eqn\xthirteen$$
where $H^{(1,2)}_{i\nu}$ denote the standard Hankel functions, and
$A(\nu)$ is a constant which is irrelevant to our discussion. The
superscript $sc$ stands for semiclassical.
Substituting these expressions in \xfourteen\ and performing the
integral over $z$ \Ref\GRAD{I. Gradsteyn and I. Ryzik, Table of Intgerals,
Series and Products (Academic Press, 1980)} one obtains
$$T_{scat}^{sc}(u,v) = B(\nu)~\mu^{{i\nu \over 2}}~
v^{- i \nu}~F(\half -i\nu,\half;1 - i\nu; -{uv \over \mu}) \eqn\xfifteen$$
where $B(\nu) = - {i {\sqrt{\pi}} \Gamma(\half - i\nu)~sinh(\pi \nu) \over
\smu \Gamma ( 1 - i\nu)}$, and $F$ stands for the hypergeometric
function.
This expression is valid throughout Region I while in Region II it is valid
for $-uv < \mu$ (note in Region II $uv$ is negative), i.e. in the region
between the horizon and the singularity. In the exterior region this
represents a wave which vanishes on the past horizon and may be written
(by using standard relations between hypergeometric functions)
in terms of a specific linear combination of right and left moving waves
at an infinite distance away from the horizon. Thus we have an incoming wave
and an outgoing wave, but nothing emerging from the past horizon - precisely
what we wanted. At the black hole singularity $uv = - \mu$ the wave is
singular. The singularity comes from the upper limit of integration
in \xfourteen. For
large $z$ the asymptotic forms of the Hankel functions are
$$H_{i\nu}^{(1)} \sim {1 \over {\sqrt{z}}}~e^{i(z - {\pi \over 2}i \nu)}~~
H_{i\nu}^{(2)} \sim {1 \over {\sqrt{z}}}~e^{-i(z - {\pi \over 2}i \nu)}
\eqn\xsixteen$$
Using this it may be seen that in \xfourteen\ the integral of products
$H_{i\nu}^{(1)}(2rz)$ $H_{i\nu}^{(1)}(2\smu z)$ and
  $H_{i\nu}^{(2)}(2zr) H_{i\nu}^{(2)}(2\smu z)$
are finite while the cross
terms lead to a logarithmic singularity at $r = \smu$. Therefore in the
semiclassical limit a part of the wave that came in from infinity and
went inside the horizon ``crashes'' into the singularity \foot{The
transform considered in \refmark\SD does not diverge at the singularity.
However, as noted above this does not describe a physical scattering
process with the correct boundary conditions.}.

\noindent{\underbar{The Exact wavefunctions}} \nextline

For non-perturbative computations of $\tV(z,t)$ we use the fermionic
field theory of the matrix model \refmark\FERM
\refmark\MORE. As explained in  \refmark\MORE, this is done as follows.
Consider the two point function
$<V(z_1,\nu)V(z_2,-\nu)>$.  The single particle tachyon wavefunction in
the standard interpretation of the matrix model is obtaining by
computing the quantity
$$ \tV(z_2,\nu) \equiv \oint {dz_1 \over 2 \pi
i}~z_1^{-i\nu -1}~<V(z_1,\nu)V(z_2,-\nu)> \eqn\xnine$$
and then performing a suitable analytic continuation to real loop lengths.
The reasoning behind this identification is that the various powers of
$z_1$ or $z_2$ denote contributions from various operators and the
above contour intgeral simply isolates the part which is the contribution
of the tachyon operator as dictated by the correct scaling dimension.(We
will restrict our attention to nonintegral $\nu$. The interpretation
for integer $\nu$ is more involved, see e.g.
\Ref\GKN{D. Gross, I, Klebanov and M. Newmann, Nucl. Phys. B350 (1991)
621; U. Danielsson and D. Gross, Nucl. Phys. B366 (1991) 3.}
\Ref\DIS{G. Moore and N. Seiberg,
Int. J. Mod. Phys. A7 (1992) 2601}
\Ref\DDMWC{S.R. Das, A. Dhar, G. Mandal and S.R. Wadia,
Mod. Phys. Lett.  A7 (1992) 937}
and related to the symmetry structure of the theory
\Ref\AJEV{J. Avan and A. Jevicki, Phys. Lett B266 (1991) 35; Phys. Lett.
B272 (1990) 17; Mod. Phys. Lett. A7 (1992) 357.}
\Ref\DDMW{S.R. Das, A. Dhar, G. Mandal and S.R. Wadia, Int. J. Mod. Phys.
A7 (1992) 5165
; Mod. Phys. Lett. A7 (1992) 71; Mod. Phys. Lett. A7 (1992) 937.}
\refmark\DIS  \Ref\POLWIN{D. Minic, J. Polchinski and Z. Yang, Nucl.
Phys. B369 (1992) 324})
It may be also simply verified that the quantity $\tV(z,\nu)$
satisfies the Wheeler de Witt equation with {\it negative} $\mu$ at the
semiclassical level.  Thus after
analytic continuation $z \rightarrow il$ (in the sense explained in
Ref[\MORE]) this does represent the physical tachyon single particle wave
function. We are, however, not interested in extracting liouville
wave functions; we will, therefore, make no such analytic continuation
\foot{We are using a definition of the nonperturbative theory which
has a potential behaving as $-x^2$ both for positive and negative $x$
\Ref\MPR{ G. Moore, M.R. Plesser
and S. Ramgoolam, Nucl. Phys. B377 (1992) 143.}. Since we do not make
any analytic continuation, the quantities we define are unambiguous
as opposed to non-perturbative S-matrices which have to be defined by
analytic continuation to real loop lengths.}

The two point function in \xnine, or rather its $\mu$-derivative may be
computed exactly in the fermionic field theory, either directly as in
\refmark\MORE or using Ward identities to relate it to the one point
function as shown in \refmark\DDMWC. The result is (after a change of
integration variables in the formulae of these papers)
$$\eqalign{{\partial \over \partial \mu}& < V(z_1,\nu) V(z_2, -\nu)>
 = {\rm Im}~\int_0^\infty {d \xi \over \mu sinh({\xi /
2 \mu})}~e^{i\xi + {i \over 2}~(z_1^2 + z_2^2)~coth~({\xi / 2 \mu})} \times
\cr & \lbrace 2 \pi~e^{{\pi \nu \over 2}}~{sinh({\nu \xi / 2 \mu}) \over
sin(\pi \nu)}~J_{i\nu}[{z_1 z_2 \over sinh({\xi / 2 \mu})}] +
 \sum_{r=1}^{\infty} {4 i^r~r~sinh(r\xi/2) \over r^2 + \nu^2}~
J_r[{z_1 z_2 \over sinh(\xi /  2 \mu)}]\rbrace}\eqn\xtwelve$$
The semiclassical expansion is obtained by expanding in powers of the
string coupling ${1 \over \mu}$
keeping $l = \smu z$ fixed. From \xtwelve\ it is clear that this
means replacing the hyperbolic functions by their power series
expansions. To the leading order it is easy to see that the answers
depend on the combinations $\smu z_1$ and $\smu z_2$, as one would expect
on the basis of the Wheeler de Witt equations.

The single particle wave functions $\tV(z,\nu)$ comes from the first term
of the
series expansion of the Bessel function $J_{i\nu}(z_1 z_2 / sinh(\xi/2))$,
which goes as $(z_1 z_2)^{i\nu}$. The result is
$$\partial_\mu \tV(z,\nu) = {\rm Im}~A(\nu)~(z/2)^{i\nu}\int_0^\infty
{d \xi \over [sinh~(\xi/2)]^{i\nu + 1}}~
e^{i[\mu \xi + \half z^2~coth~{\xi \over 2}]}~sinh(\nu \xi/2) \eqn\xtwelvea$$
where $A(\nu) = {2 \pi e^{\pi \nu \over 2} \over sinh(\pi \nu)~\Gamma
(1 + i\nu)}$. The integration over $\xi$ may be performed
in terms of Whittaker functions \refmark\MORE. The result is
$$\partial_\mu \tV (z,\nu) = {\rm Im}~B(\nu){1 \over z}
[\Gamma (\half - i\mu)~  W_{i(\mu + {\nu \over 2}),{i\nu \over 2}}(-iz^2)
-
\Gamma (\half - i\mu + i{\nu \over 2})~
W_{i(\mu - {\nu \over 2}),{i\nu \over 2}}(-iz^2)] \eqn\xseventeen$$
where $W_{\lambda,\kappa}$ stands for the Whittaker function.
For zero energy $\nu = 0$
this is the $\mu$- derivative of the vacuum expectation value of the
operator $V(z,t)$ as expected. This takes the simple form
$$\tV(z,0) = {\rm Im}~[e^{-3 \pi i \over 4}~\Gamma(\half - i\mu)~{1 \over z}
{}~W_{i\mu,0} (-iz^2)] \eqn\xeighteen$$

Several properties of these exact expressions for $\tV(z,\nu)$ are worth
noting and will be useful in a moment. First, to take the semiclassical
limit one has to remember that $l = \smu z$ has to be kept fixed while
$\smu \rightarrow \infty$.
Using standard asymptotic forms of Whittaker functions
\Ref\ABSTEG{M. Abramowitz and I. Stegun, {\it Handbook of Mathematical
functions} (Cambridge, 1962)} it may be checked that the above expression
reduce to the semiclassical expressions involving Hankel functions in
this limit. Secondly, {\it for small values of $z$ these wavefunctions
reduce to the semiclassical wavefunctions in the corresponding limit}.
This may be seen from \xtwelvea\ from where it is clear that for small
$z$ only large values of $coth~{\xi \over 2}$, or
only small values of $\xi$ contribute. In that case one can replace
the hyperbolic functions by their power series expansions. But this is
precisely what yields the semiclassical limit. Alternatively one can directly
take the result \xseventeen\ and use the small-$z$ behaviour of the
Whittacker functions
\refmark\ABSTEG and verify that one indeed recovers the same
answer as one obtains from the small $z$ behaviour of the Hankel functions
involved in the semiclassical wave function.
This is simply a reflection of the fact that the coupling constant in
the string field theory of the matrix model goes rapidly to zero away
from the wall. In our case this region corresponds to the region of
small $z$.

The quantity $\partial_\mu T_{scat}(u,v)$ can be now calculated by inserting
\xseventeen\ in \xfourteen.
 The first point to note about the reaulting
expressions for the exact $T_{scat}(u,v)$ is that for large values of
$\vert uv \vert$, i.e. for large values of $r$ the exact expressions
agree with the semiclassical expressions in this region. This follows
from the large $r$ behavior of the Bessel functions in \xfourteen.
However, we have just argued
that for small $z$ the exact wavefunctions $\tV(z,\nu)$ approach their
semiclassical estimates. Thus for large $r$ the black hole wavefunctions
also agree with their semiclassical answers. This fact shows that in the
asymptotic region of the black hole the coupling of the theory is indeed
weak. By the same token we expect that near the black hole there will be
significant deviations from the semiclassical behavior.

We saw in that in the semiclassical limit the logarithmic singualrity of
the scattering wavefunction $T_{scat}$ came from the upper limit of
integration, i.e. from large $z$. This came about from a competition
between the exponentials coming from the matrix model wavefunctions and the
Bessel functions in \xfourteen. However it
is precisely in the large $z$ region that the matrix model wavefunctions
differ most from the semiclassical answers. In fact, the asymptotic
behavior of the Whittaker function is given by
$$W_{\lambda ,\kappa} (-i z^2) \sim z^{2\lambda}~e^{{iz^2 \over 2}}
\eqn\xnineteen$$
and there is no competition between this and the $e^{2izr}$ coming from the
Hankel functions. The result is in fact {\bf finite at the singularity}.

It is important to note that the scattering wavefunction remains
infinite at the location of the black hole singularity to {\it all orders
in the semiclassical expansion}. This is similar to the result of
\refmark\DMWBH about the value of the one point function. The
smearing of the behavior of the wavefunction near the singularity is
a genuinely nonperturbative effect.

\vskip 1.4cm

\noindent{\underbar{Discussion}}

What does this result mean ? Our matrix model black hole connection
shows that one can obtain wavefunctions (and presumably the S-matrix)
of tachyons moving in a black hole background at the semiclassical level
by evaluating suitable quantities in the matrix model. Given this
connection we evaluate the {\it same} quantities non-perturbatively
in the matrix model. Our interpretation of these non-perturbative
answers is that we are evaluating correlations or wavefunctions in
{\it some} (possibly nonlocal) theory of a scalar field $T(u,v)$ whose
semiclassical behavior is identical to that of tachyons in a black hole
background.

We found that while the exact wavefunctions approach the semiclassical
wavefunctions in the asymptotic region, they depart significantly inside
the black hole, which is therefore a region of strong coupling.
In this region it is not very meaningful to expand around
the standard classical ground state solution and interpret the
resulting fluctuations as ``particles''. Nevertheless $\tT_{scat} (u,v)$
can be correctly interpreted as single particle tachyon wavefunctions
in the aymptotic regions and hence it is meaningful to talk of a
scattering process.

As emphasized in \refmark\DVV the region ``beyond the singularity'',
i.e. the portion of Region II with $-uv > \mu$ has to be taken
seriously for the string theory black hole.
Since the exact answer for $\tT_{scat}(u,v)$ is nonsingular it is now
meaningful to continue it across the singularity and ask what happens
``beyond the singularity''. For large $r$ in Region II we can either
look at the small $z$ behavior of the Whittaker functions
\refmark\ABSTEG or simply
look at the semiclassical answer. This can be easily computed (the
answer \xfifteen\ is not valid here since $-uv > \mu$)
$$T_{scat}^{(\nu)}(u,v) \sim {1 \over r}~
({r \over \smu})^{i\nu}~e^{i \nu \theta} \eqn\xtwenty$$
which shows that the wavefunction has support on {\it one} of the null
infinities (corresponding to large values of $v$). The wave simply
lands up in the asymptotic region beyond the singularity.

To really see whether we are describing the behavior
of tachyons of {\it two dimensional string theory} in a black hole
background, one has to compute $S$-matrices
and compare them with continuum $S$-matrices, e.g. those obtained from
the coset model. After all, the identification of quantities
in the matrix model with tachyon fields in a liouville background
can be considered as correct because the tree level S-matrix
computed by matrix model methods
\Ref\MSW{G. Mandal, A. Sengupta and S.R. Wadia, Mod. Phys. Lett. A6
(1991) 1465;  K. Demeterfi, A. Jevicki and J.P. Rodrigues, Nucl. Phys. B362
(1991) 173 and Nucl. Phys. B365 (1991) 199; D. Gross and I. Klebanov,
Nucl. Phys. B359 (1991) 3;
J. Polchinski, Nucl. Phys. B362
(1991) 125; G. Moore, Nucl. Phys. B368 (1992) 557.}
agreed with those obtained from the continuum
\Ref\DIFRKUT{P. Di Francesco and D. Kutasov, Phys. Lett. B261 (1991) 385.}.
Without a similar continuum calculation in a black hole background
the relationship between the theory of
tachyon fields which result from the transform discussed above and
string theory is unclear. Nevertheless our results show that in
this exactly solvable
theory of tachyons nonperturbative effects may drastically alter the
behavior near the singularity.

In the standard interpretation of the matrix model it is
troublesome to assign a clear spacetime interpretation in the
strongly coupled region \refmark\DMW \refmark\MORE. For example
there is no relativistic invariance in this region. In our case
this implies that the region inside the black hole does not have a
clear relativistic space time interpretation.
This is, in fact, expected in a string theory.
As noted above, the fact
that the theory is weakly coupled far from the black hole allows us
to meaningfully desribe scattering processes involving tachyons.
However this makes the absence of singularities of the exact
wavefunction difficult to understand.
Assuming that our complicated field theory of massless ``tachyons''
is indeed a {\it string} theory, the only statement we can make
is that the string does not ``see'' any singularity even though
it perceives a singularity at the semiclassical level. Since the
graviton and dilaton fields are not explicitly present in the
formalism it is confusing to ask whether the singularity itself
has vanished.

Finally, it would be very interesting if one can understand
time dependent backgrounds in string theory,
like formation of black holes, in terms of the matrix model. Since there is no
explicit field for the graviton or the dilaton it would not be possible
to see the black hole ``evaporate away'', but its effects on the
scattering of tachyons can still be studied.

\vskip 3.0cm
\centerline{\underbar{Acknowledgements}}

I would like to thank A. Dhar, J. Harvey, D. Kutasov,
G. Mandal, E. Martinec,
Y. Nambu, S. Trivedi and S. Wadia for discussions. This work was
supported in part by the U.S. Department of Energy grant No.
DEFG02-90ER-40560, the NSF grant No. PHY-91-23780 and a University
of Chicago grant No. 2-64664. I would like to thank P. Freund and
all other members of the Theoretical Particle Physics group of
the Enrico Fermi Institute for their hospitality.

\endpage
\refout

\end